\documentclass[preprint,showpacs,amsmath,amssymb]{revtex4}

\usepackage{graphicx}
\usepackage{dcolumn}
\usepackage{bm}

\begin{document}

\title{Field-Dependent Hall Effect in Single Crystal Heavy Fermion YbAgGe below 1K}

\author{S.L. Bud'ko$^\ast$, V.Zapf$^\dagger$, E. Morosan$^\ast$, and P.C. Canfield$^\ast$}
\affiliation{$^\ast$Ames Laboratory US DOE and Department of Physics and Astronomy, Iowa State University, Ames, IA 50011, USA\\
$^\dagger$MS E536, National High Magnetic Field Laboratory - Los Alamos National Laboratory, Los Alamos, NM 87545,
USA}

\date{\today}

\begin{abstract}
We report the results of a low temperature ($T \geq 50$ mK) and high field ($H \leq 180$ kOe) study of the Hall
resistivity in single crystals of YbAgGe, a heavy fermion compound that demonstrates field-induced
non-Fermi-liquid behavior near its field-induced quantum critical point. Distinct features in the anisotropic,
field-dependent Hall resistivity sharpen on cooling down and at the base temperature are close to the respective
critical fields for the field-induced quantum critical point. The field range of the non-Fermi-liquid region
decreases on cooling but remains finite at the base temperature with no indication of its conversion to a point
for $T \to 0$. At the base temperature, the functional form of the field-dependent Hall coefficient is field
direction dependent and complex beyond existing simple models thus reflecting the multi-component Fermi surface of
the material and its non-trivial modification at the quantum critical point.
\end{abstract}


\maketitle

\section{Introduction}

YbAgGe was recently recognized to be one the few stoichiometric Yb-based heavy fermion compounds that demonstrate
field-induced non-Fermi-liquid (NFL) behavior as evidenced by thermodynamic and transport measurements in an
applied magnetic field \cite{bud04a}. In zero applied field YbAgGe has two (small magnetic moment) magnetic
ordering transitions, at $\approx 1$ K and $\approx 0.65$ K (first order) \cite{bud04a,mor04a,ume04a}. The
magnetic ordering temperatures are well separated from the Kondo temperature ($T _K \sim 25$ K) that is well below
estimated CEF splitting ($T_{CEF} \sim 60-100$ K) \cite{kat04a,mat04a}. The critical field needed to reach the
quantum critical point (QCP) in YbAgGe is moderate and anisotropic ($H_c^{ab} \approx$ 45 kOe, $H_c^c \approx$ 80
kOe) \cite{bud04a,ume04a}. Hall effect measurements through the QCP have been suggested \cite{col01a} to be one of
the key experiments to distinguish between two general descriptions of an antiferromagnetic QCP: (1)
spin-density-wave and (2) composite heavy fermion (HF) scenarios. Initial (down to 0.4 K) Hall effect measurements
on YbAgGe \cite{bud05a} show clear local maxima/minima (for $H \| ab$/$H \| c$, respectively) in the
field-dependent Hall resistivity, $\rho_H(H)$, that occur at values that approach the respective critical fields
as $T \to 0$. The feature in $\rho_H(H)$ establishes a new, distinct, line in the anisotropic $H - T$ phase
diagrams of YbAgGe \cite{bud05a}. A similar, additional, Hall-effect line was detected in YbRh$_2$Si$_2$
\cite{pas04a}, the other Yb-based compound that is perceived to be a well established example of a material with a
field-induced QCP \cite {geg03a}. For YbRh$_2$Si$_2$, the gradual change of the Hall coefficient, $R_H$, in an
applied field was interpreted \cite{pas04a} as extrapolating to a sudden jump at the QCP in the zero temperature
limit and hence suggesting that the composite HF scenario \cite{col01a} is realized in this material. As distinct
from YbRh$_2$Si$_2$, the field-dependent Hall coefficient in YbAgGe, even above 0.4 K (the base temperature in our
previous study), had a rather complex, albeit fairly sharp, feature at $H_c$, clearly different for the
field-induced QCP for $H \| ab$ and $H \| c$ \cite{bud05a}. Extension of the temperature range of the Hall effect
measurements down to 50 mK in temperature allows us to observe the evolution of the field-dependent Hall
coefficient much closer to the $T = 0$ QCP. In this work we expanded the probed $H - T$ space down to $T/T_N \sim
0.05$ and up to $H/H_{QCP} \sim 2 - 4$ that exceeds considerably our previous capabilities as well as, in relative
($T/T_N, H/H_{QCP}$) terms, the phase space accessed in Hall measurements on YbRh$_2$Si$_2$ \cite{pas04a}.

\section{Experimental}

YbAgGe single crystals in the form of clean, hexagonal-cross-section rods of several mm length and up to 1 mm$^2$
cross section were grown from high temperature ternary solutions rich in Ag and Ge (see \cite{mor04a} for details
of the samples' growth). Their structure and full site-occupancy without detectable site-disorder were confirmed
by single crystal X-ray diffraction \cite{moz04a}. The four-probe Hall resistivity measurements $\rho_H(H)|_T$
were performed on two pairs of the samples: (i)$H\|[120]$, $I\|[001]$ with the Hall voltage, $V_H$, measured along
$[100]$; (ii)$H\|[001]$, $I\|[100]$ with $V_H$ measured along $[120]$. One sample in each pair was exactly the
same sample as studied in Ref. \onlinecite{bud05a}. The measurements were performed at temperatures down to 50 mK
and magnetic fields up to 180 kOe using a $^3$He-$^4$He dilution refrigerator in a 200 kOe Oxford Instruments
superconducting magnet system at the National High Magnetic Field Laboratory in Los Alamos. The samples were
immersed in the $^3$He-$^3$He mixture together with a field-calibrated RuO$_2$ thermometer, thus providing
excellent thermalization and allowing for use of higher excitation currents to achieve better signal to noise
ratio. The Hall resistance was measured with a Linear Research LR-700 ac resistance bridge with excitation
currents of 1-3 mA (no additional sample heating was detected at the base temperature and above at these current
values). For most of the runs the protocol of the measurements was the following: the temperature was stabilized
at the desired value with the 180 kOe (-180 kOe) field applied, then the measurements were taken while the field
was swept at 2 kOe/min rate through zero to -180 kOe (180 kOe). Magnetic flux jumps in the 200 kOe superconducting
magnet drastically increase the noise in the data taken in applied field below $\sim 20$ kOe, so only data for $H
\geq 20$ kOe will be presented. To eliminate the effect of inevitable (small) misalignment of the voltage
contacts, the Hall measurements were taken for two opposite directions of the applied field, $H$ and $-H$, and the
odd component, $(\rho_H(H)-\rho_H(-H))/2$ was taken as the Hall resistivity, $\rho_H(H)$. Since, within the error
bars of the measured geometry of the samples and contact positions, both samples in each pair yielded the same
results, for ease of comparison in the following we will present the data for the samples used in Ref.
\onlinecite{bud05a}.

\begin{figure}
\begin{center}
\includegraphics[angle=0,width=80mm]{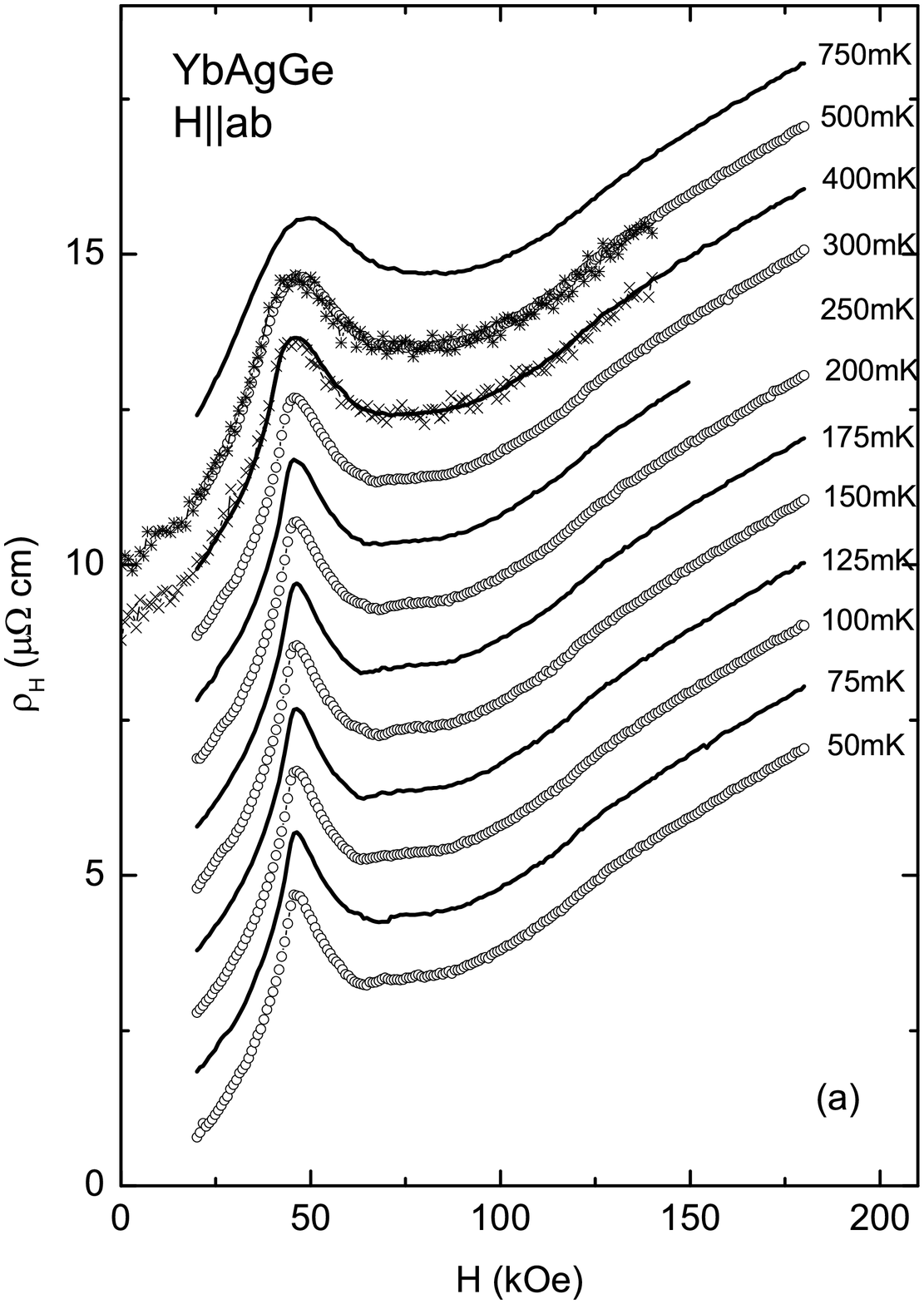}
\includegraphics[angle=0,width=80mm]{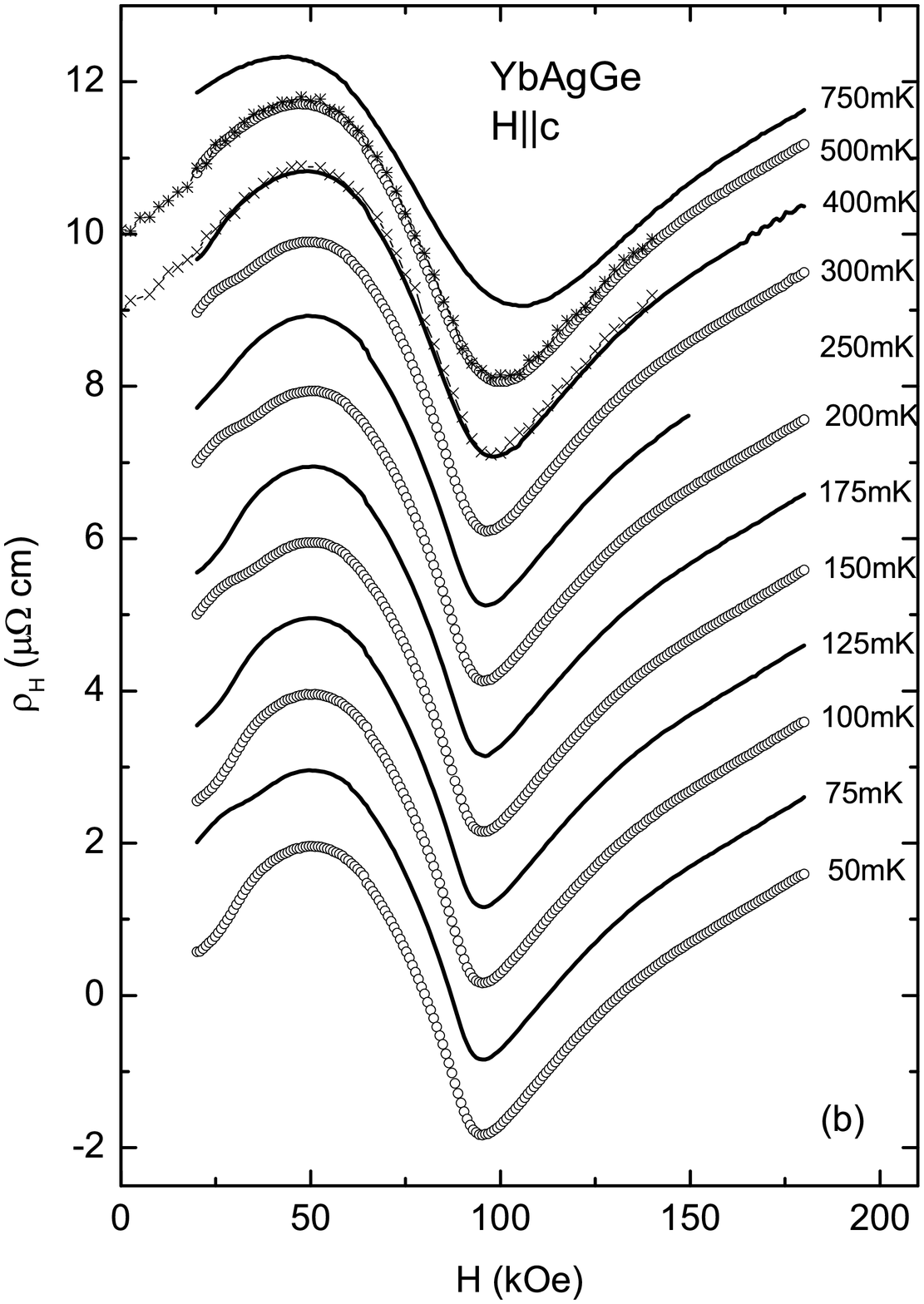}
\end{center}
\caption{Field-dependent Hall resistivity of YbAgGe ((a) - $H \| ab$, (b) - $H \|c$) measured at different
temperatures. The curves, except for $T$ = 50 mK, are shifted by $1 \mu \Omega$ cm increments for clarity. For 400
mK and 500 mK, data from our initial measurements \cite{bud05a} are shown by $\times$ and $\ast$ for comparison.
Note: for (b) the feature located near $H = 20-30$ kOe is an artifact of the measurements protocol and hysteresis
in magnetoresistivity at the lower transition (see text).}\label{rhoH}
\end{figure}

The protocol of the measurements adopted in this work, whereas time-conserving, resulted in an artifact
(originating from the hysteretic component of the magnetoresistance at the lower magnetic transition
\cite{bud04a,mor04a,ume04a}, which is not eliminated completely through the Hall resistivity calculations
described above in the measurements protocol used for the most of the measurements) seen in the low temperature
$\rho_H(H)$ data for $H \| c$ at approximately 20-30 kOe (see Fig. \ref{rhoH}(b) below). Whereas we kept this
artifact in $\rho_H(H)$ data \ref{rhoH}(b) for illustrative purpose, subsequent data for $H \| c$ were truncated
to $\sim 30$ kOe. For $H \| ab$ a similar feature occurs below $\sim 20$ kOe, in the region of magnetic flux jumps
noise.

\section{Results and Discussion}

Field-dependent Hall resistivities for $H \| ab$ and $H \| c$ at temperatures between 50 and 750 mK are shown in
Fig. \ref{rhoH}. In the overlapping temperature region, the current data are consistent with that reported in Ref.
\onlinecite{bud05a}. On cooling down, the main feature in the $\rho_H(H)$ sharpens for both orientations of the
applied field. For $H \| ab$, in addition to sharpening, a plateau in $\rho_H(H)$ emerges from the data for $T <
400$ mK in the approximate range of fields 60 kOe $< H <$ 90 kOe.

\begin{figure}
\begin{center}
\includegraphics[angle=0,width=80mm]{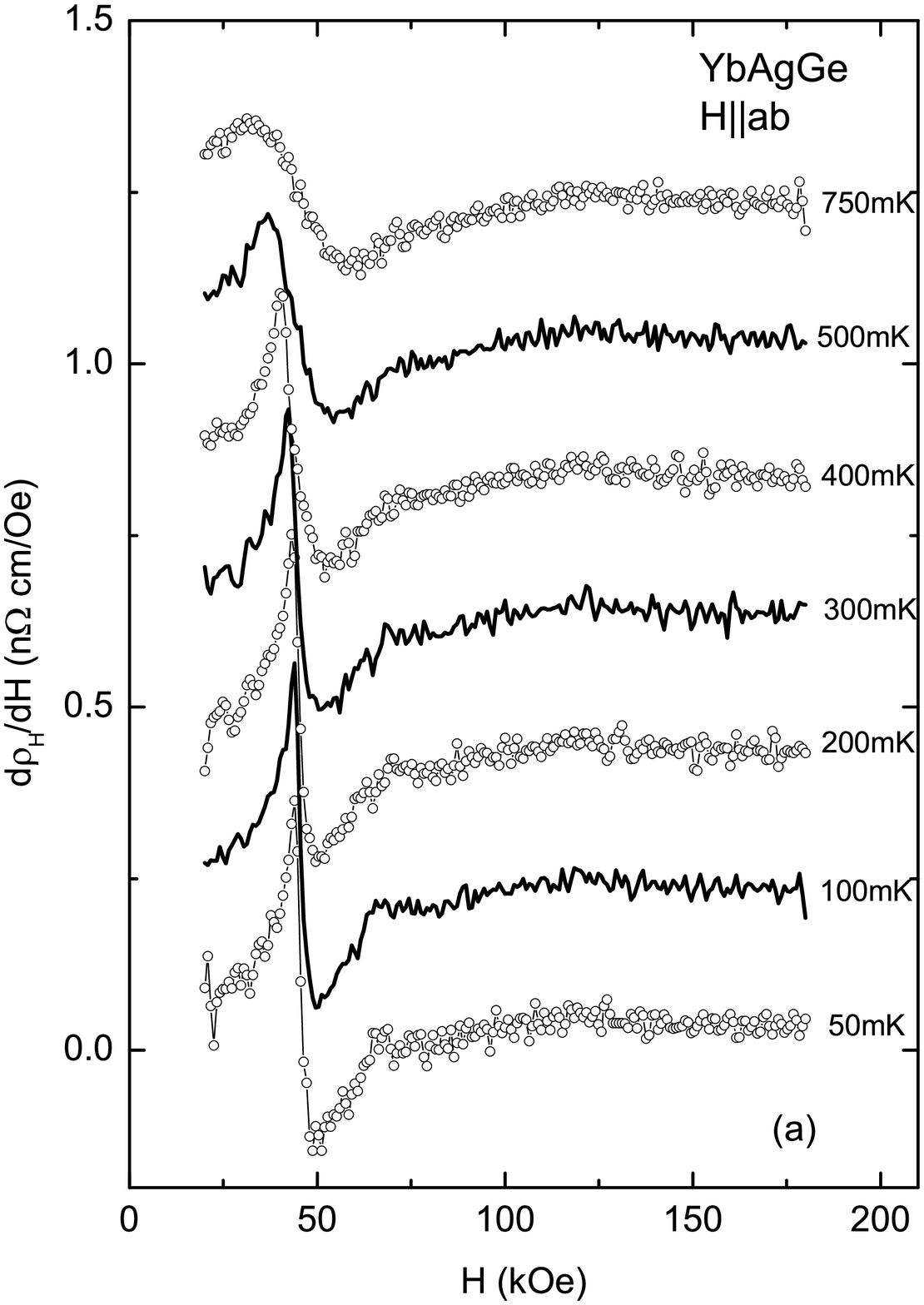}
\includegraphics[angle=0,width=80mm]{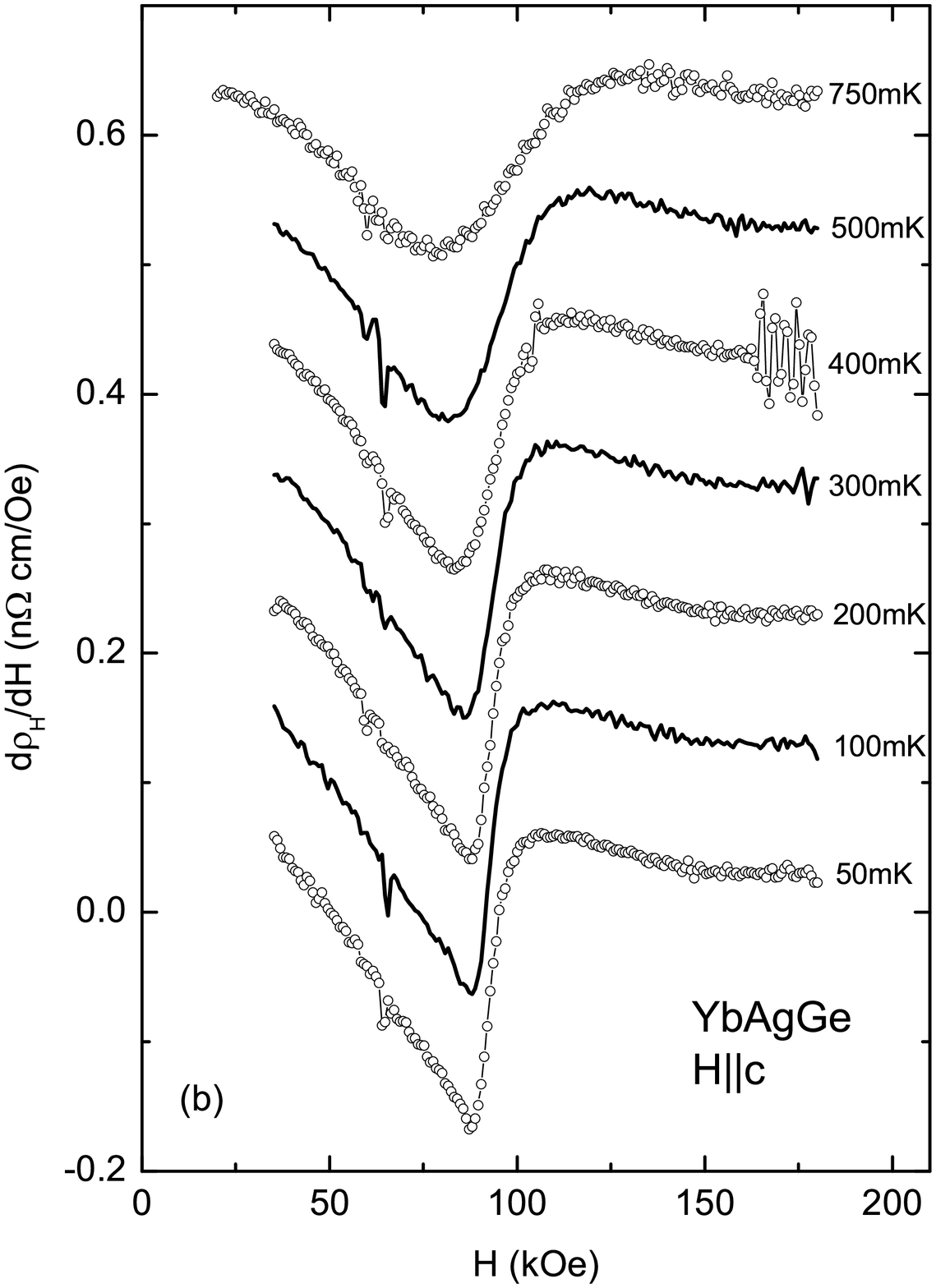}
\end{center}
\caption{Field-dependent Hall coefficient of YbAgGe ((a) - $H \| ab$, (b) - $H \|c$), defined as $R_H = d \rho_H/d
H$, measured at different temperatures. The curves, except for $T$ = 50 mK, are shifted by (a) 0.2 n$\Omega$ cm/Oe
and (b) 0.1 n$\Omega$ cm/Oe increments for clarity. Low field data in (b) (except for 750 mK, above the lover
magnetic transition) are truncated below $\sim 30$ kOe (see text).}\label{derrhoH}
\end{figure}

\begin{figure}
\begin{center}
\includegraphics[angle=0,width=80mm]{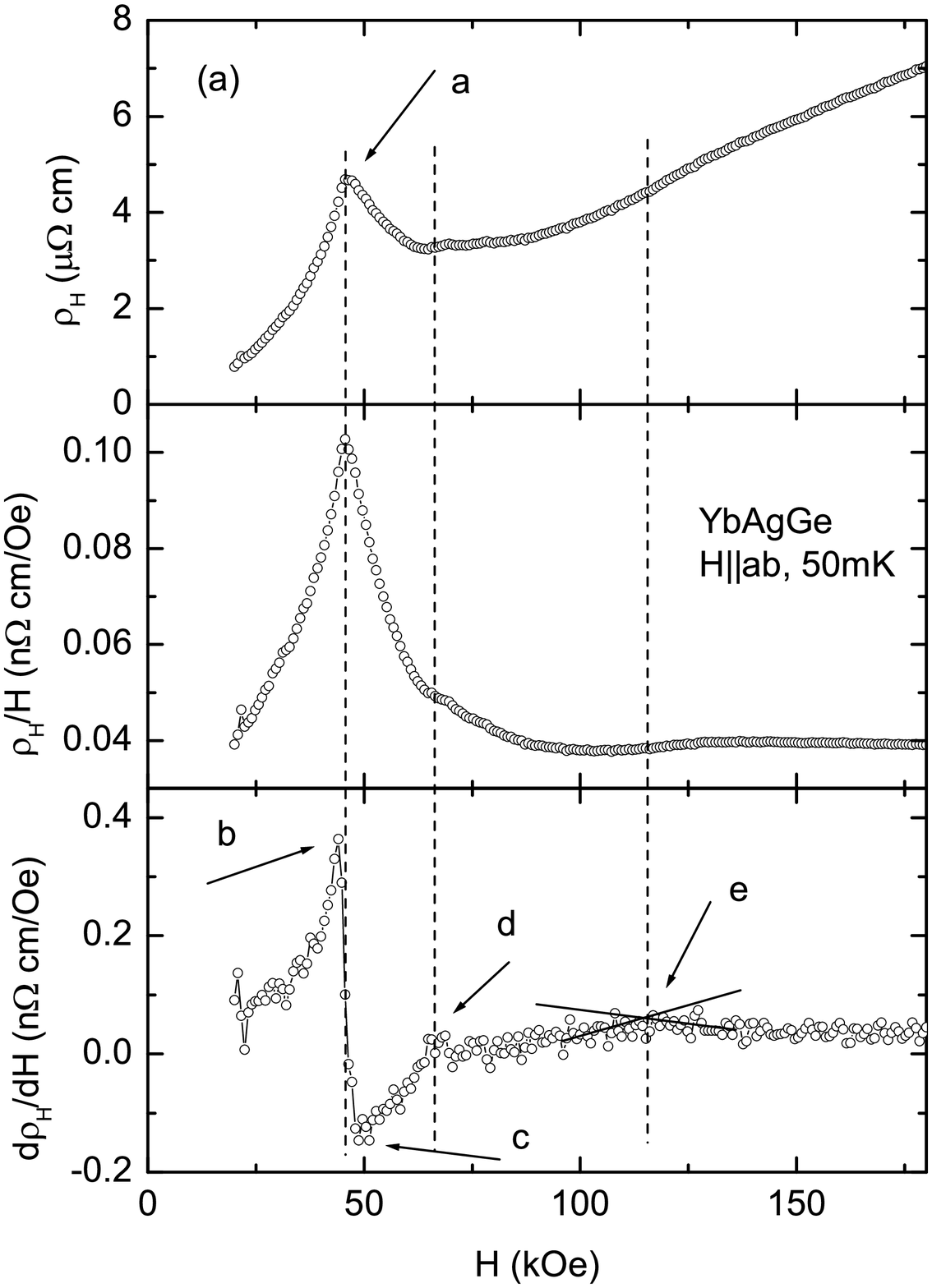}
\includegraphics[angle=0,width=80mm]{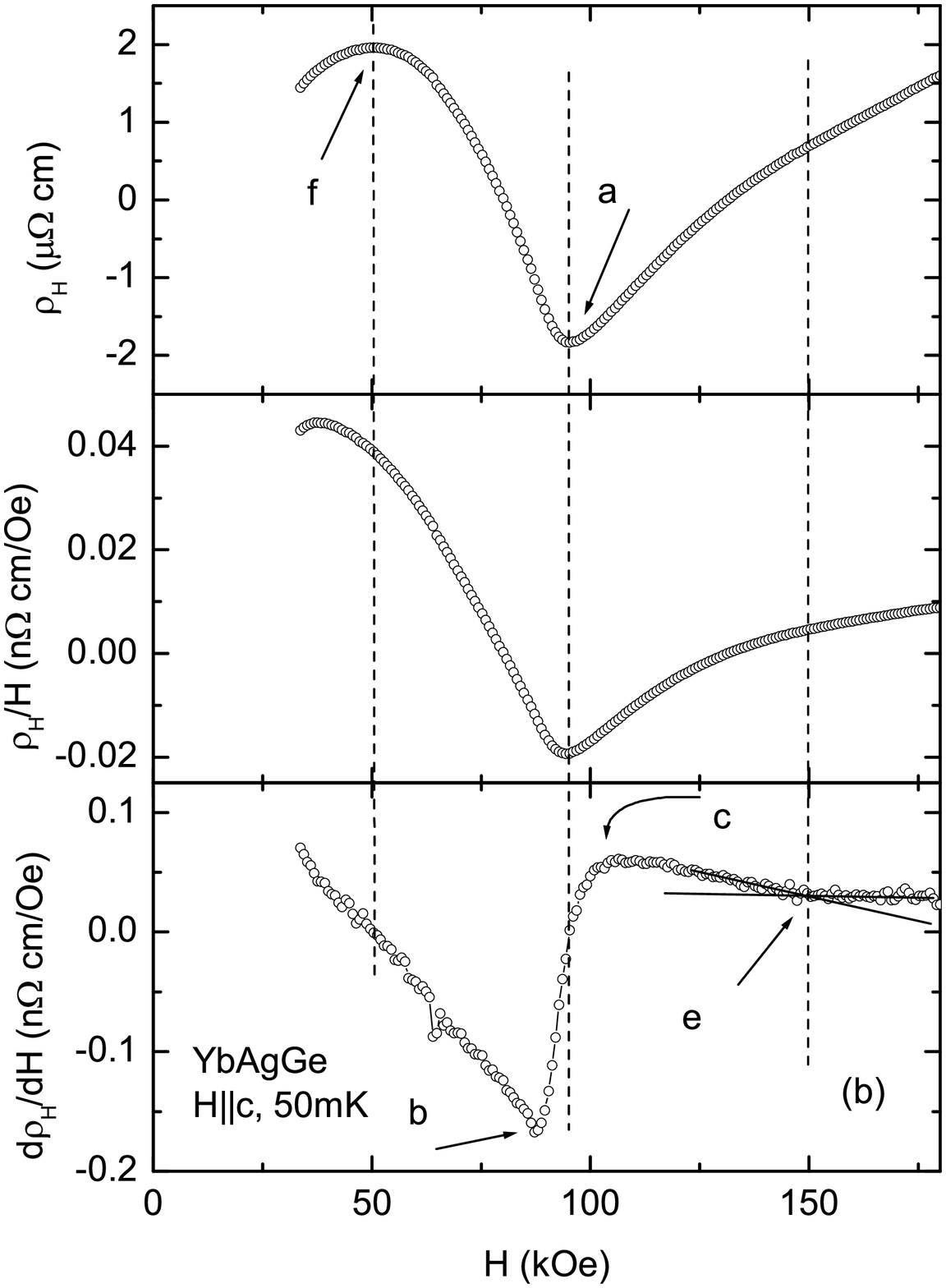}
\end{center}
\caption{Field-dependent Hall resistivity ($\rho_H$), Hall resistivity divided by field ($\rho_H/H$) and Hall
coefficient ($d \rho_H/d H$) of YbAgGe ((a) - $H \| ab$, (b) - $H \|c$) at $T = 50$ mK. Arrows and letters mark
different features of the curves (see text for the discussion).}\label{deffeat}
\end{figure}

Representative field-dependent Hall coefficients (calculated as $d \rho_H/d H$ from the sub-set of the data in
Fig. \ref{rhoH}) are shown in Fig. \ref{derrhoH}. As in the case of Hall resistivities, the features associated
with the proximity to field induced QCP are sharper and better-defined at lower temperatures. In comparison to the
initial study \cite{bud05a}, lower temperatures, higher fields and better signal-to-noise ratio of the current
data allow us to consider more than one feature in the Hall resistivity and Hall coefficient (Fig. \ref{deffeat}).
The least ambiguous feature is a sharp maximum/minimum ($H \| ab$/$H \|c$) in the anisotropic Hall resistivity
(a). The two extrema in $d \rho_H/d H$, (b) and (c), are associated with the main feature in $\rho_H$ and are, in
a broad sense, a measure of the width of the respective feature. Clear breaks of the slope ((d), (e) for $H \|
ab$, (e) for $H \| c$) are seen at higher fields. It is noteworthy that for $H^{ab} > H_e$ and $H^c > H_e$ the
Hall coefficient, $R_H = d \rho_H/d H$, is practically field-independent, suggesting these fields to be a caliper
of the material entering into Fermi-liquid state and as such approximately defining the coherence line on the $H -
T$ phase diagram (alternatively plotted for this material in Refs. \onlinecite{bud04a,bud05a} as the temperature
up to which the resistivity obeys $\rho(T) = \rho_0 + AT^2$ law). Finally, for $H \| c$ there is a relatively
broad (even at the base temperature) maximum (f) in the Hall resistivity, however, since there is no corresponding
feature in $d\rho_H/dH$ (lower panel of Fig. \ref{deffeat}(b)) it is not so likely that this maximum corresponds
to a phase boundary or crossover.

\begin{figure}
\begin{center}
\includegraphics[angle=0,width=80mm]{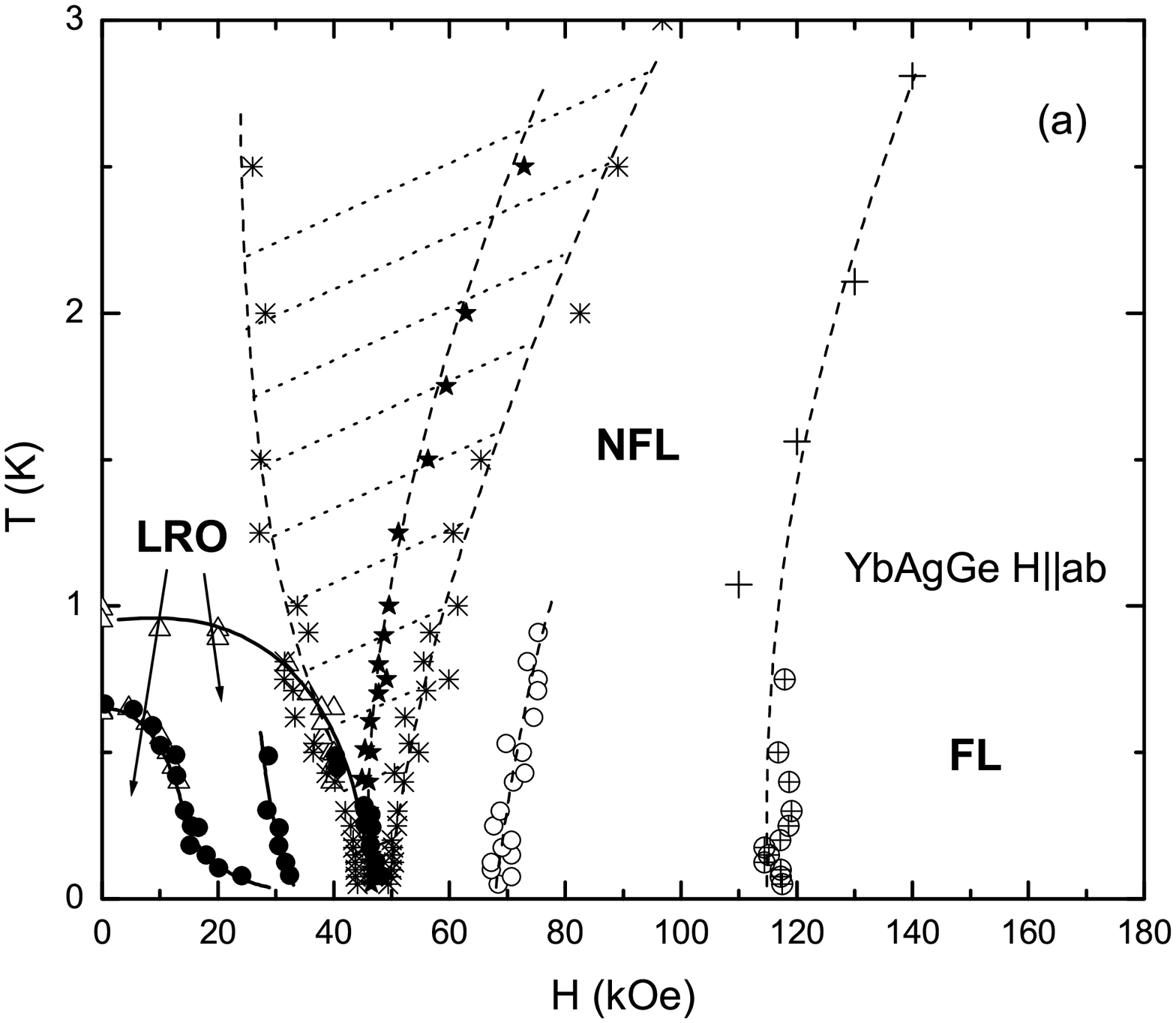}
\includegraphics[angle=0,width=80mm]{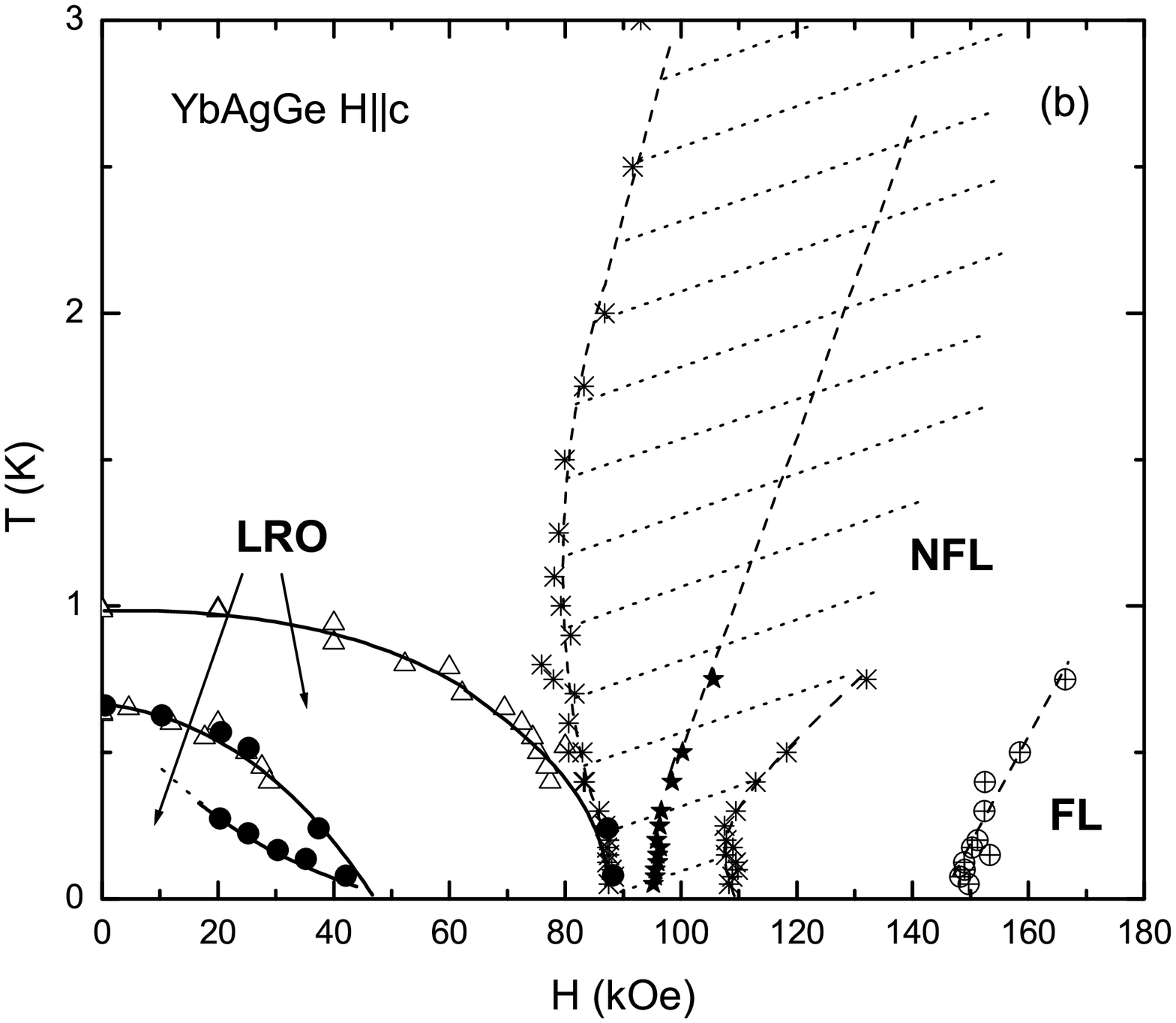}
\end{center}
\caption{Composite anisotropic $H - T$ phase diagrams for YbAgGe: (a) $H \| ab$, (b) $H \| c$. Symbols:
$\triangle$ - magnetic transitions from thermodynamic and magneto-transport measurements in Ref.
\onlinecite{bud04a}; $\bullet$ - magnetic transitions from magneto-transport measurements in Ref.
\onlinecite{nik05a}; solid lines - tentative magnetic phase boundaries; $+$ - coherence line defined as a high
temperature limit of the region of $\rho(T) = \rho_0 + AT^2$ resistivity behavior \cite{bud04a}; $\bigstar$ - Hall
line defined from the maximum/minimum ($H \| ab$/$H \| c$) of the Hall resistivity ((a) in Fig. \ref{deffeat});
$\ast$ - lines defined from the maximum and minimum in $d \rho_H/d H$ ((b) and (c) in Fig. \ref{deffeat}), area
between these lines is accentuated by hatching; $\circ, \oplus$ in panel (a) and $\oplus$ in panel (b) - point
corresponding to high field break of the slope in $d \rho_H/d H$ ((d) and (e) in Fig. \ref{deffeat}(a) and (d) in
Fig. \ref{deffeat}(b)). Dashed lines are guide for the eye.}\label{PD}
\end{figure}

Based on the measurements and salient features discussed above, considered together with the thermodynamic and
transport data in an applied field at temperatures down to 400 mK \cite{bud04a}, the magnetotransport measurements
down to 70 mK \cite{nik05a} and our earlier Hall measurements down to 400 mK \cite{bud05a}, we can construct
anisotropic low-temperature $H - T$ phase diagrams for YbAgGe (Fig. \ref{PD}). For a discussion related to the
complexity of the magnetically ordered phases, we refer the reader to the original works
\cite{bud04a,mor04a,ume04a,bud05a,nik05a}. The Hall line defined from the maximum/minimum ($H \| ab$/$H \| c$) of
the Hall resistivity at low temperatures approaches the QCP. The width of the Hall anomaly related to the QCP
(defined here as the region between two relevant extrema, (b) and (c) in Fig. \ref{deffeat}, in $d \rho_H/d H$)
decreases on cooling down and reaches $\approx 0.2 H_{crit}$ at the base temperature, being much narrower, in
relative units, than for YbRh$_2$Si$_2$ \cite{pas04a}. For both orientations of the applied field the Hall anomaly
in YbAgGe can be followed up to $2 - 3$ K, far beyond the ordering temperature in zero field. At low temperatures,
the lower-field boundary of the Hall anomaly initially follows the magnetic phase boundary (Fig. \ref{PD}) and
then departs from this boundary and continues singly to higher temperatures.

Hall measurements in an extended $H - T$ range suggest an existence of additional boundaries on the phase
diagrams. For $H \| ab$ (Fig. \ref{PD}(a)), almost temperature-independent lines at $\sim 70$ kOe and $\sim 120$
kOe are suggested by following two high-field kinks in $d \rho_H/d H$. Above $\sim 1$ K there features are smeared
out and are difficult to follow further. The Hall coefficient becomes field-independent above the 120 kOe line.
The line starting at $\sim 120$ kOe can be easily extended to accommodate the coherence temperature points defined
from the magnetotransport $\rho(T)|_H$ data in Ref. \onlinecite{bud04a} into a common phase boundary. For $H \| c$
(Fig. \ref{PD}(b)) $d \rho_H/d H \approx const$ for fields above the line that corresponding to $\sim 150$ kOe at
the base temperature. Similarly to the case of $H \| ab$ above, this line can be used to re-define the coherence
line for $H \| c$ and shift it to higher fields compared to its tentative position inferred from the limited
$\rho(T)|_H$ data \cite{bud04a} taken in significantly smaller $H - T$ domain. It is worth noting that Figures
\ref{PD}(a) and \ref{PD}(b) indicate that there appears to be a finite range of fields, even as $T \to 0$, over
which NFL appears. This is rather remarcable and needsw further theoretical, as well as experimental, study.

Whereas the general features of the phase diagram discussed so far and their evolution with temperature are, at a
gross level, similar for both orientations of the applied field, the functional form of the field-dependent Hall
coefficient (Fig. \ref{derrhoH}), differs between $H \| ab$ and $H \| c$, and is disparate to the theoretically
suggested options \cite{col01a} and the observations in YbRh$_2$Si$_2$ \cite{pas04a}. Whereas the level of
similarity between the field-induced NFL behavior in YbAgGe and YbRh$_2$Si$_2$ is open for debate, it is of note
that the Hall resistivity (or Hall coefficient) in real materials has a complex dependence on the details of the
electronic structure (see {\it e.g.} Refs. \onlinecite{liv99a,hal05a}). Even in a simpler (but somewhat related)
case of the electronic topological transition (ETT) \cite{var89a,bla94a}, in the absence of strong electronic
correlations, an accurate prediction (based on band-structure calculations) of the change of the Hall coefficient
as a function of the parameter controlling the ETT may be an intricate task \cite{liv02b}. With this in mind, the
differences in functional behavior of the Hall coefficient for two orientations of the applied field in YbAgGe is
probably due to the details of the complex anisotropic Fermi surface of this material, while the distinct feature
for each direction of the applied field is associated with the field-induced QCP in this material.

\section{Summary}

Distinct features in the Hall resistivity of YbAgGe for two orientations of the applied magnetic field, $H \| ab$
and  $H \| c$, were followed down to $T = 50$ mK. The features sharpen on cooling down and at the base temperature
are close to the respective critical fields for the field induced QCP. New lines are being suggested for the
composite $H - T$ phase diagrams. The non-Fermi-liquid part of the $H - T$ phase diagram apparently remains finite
in the limit of $T \to 0$ suggesting that the topology of the phase diagram in the vicinity of the field-induced
QCP in YbAgGe is different from most commonly acknowledged cases.

\begin{acknowledgments}
Ames Laboratory is operated for the U.S. Department of Energy by Iowa State University under Contract No.
W-7405-Eng.-82. This work was supported by the Director for Energy Research, Office of Basic Energy Sciences. Work
at the NHMFL - Los Alamos was performed under the auspices of the NSF, the state of Florida and the U.S.
Department of Energy.
\end{acknowledgments}

\clearpage

\clearpage

\clearpage

\clearpage

\end{document}